\documentstyle[11pt]{article}
\newfont{\gothic}{eufm10 scaled\magstep 1}

\def\bigstar{\;\bigcirc\kern-0.9em\star\;}
\begin{document}

{\Large \centerline{NAMBU BRACKETS WITH CONSTRAINT FUNCTIONALS}

\centerline{Adnan TE\u{G}MEN}

Department of Physics, University of Ankara,
06100 Tando\u{g}an Ankara TURKEY\\
\phantom{.} \qquad\qquad{\sl tegmen@science.ankara.edu.tr}

\begin{abstract}
If a Hamiltonian dynamical system with $n$ degrees of freedom
admits $m$ constants of motion more than $2n-1$, then there exist
some functional relations between the constants of motion. Among
these relations the number of functionally independent ones are
$s=m-(2n-1)$. It is shown that for such a system in which the
constants of motion constitute a polynomial algebra closing in
Poisson bracket, the Nambu brackets can be written in terms of
these $s$ constraint functionals. The exemplification is very rich
and several of them are analyzed in the text.
\end {abstract}

\vskip 0.5cm

\section{Introduction}

The concept of generalized Hamiltonian dynamics arouse in 1973
with an
article by Y. Nambu. \cite{Nambu} In his proposal, Nambu employed an $N$%
-ary bracket, generically called Nambu bracket (NB), to describe
the time evolution of the dynamical system in $N$-dimensional
$(ND)$ phase space. His bracket includes $N-1$ functionally
independent constants of motion, the so-called generalized
Hamiltonians. As an illustrative example, Nambu considered the
Euler equations of free rigid body for a $3D$ phase space and this
was the only example given. Finding examples in higher
odd-dimensions is still very tedious matter.

In Nambu formalism, dynamical systems produce inevitably a
nontrivial normalization factor $C$ at least when $N$ is an even
integer grater than three. \cite{Chatterjee} In words, in order to
get the correct Hamiltonian dynamics the NBs must be normalized
properly. The explicit general form of $C$ has been derived in
detail for superintegrable systems. \cite{Tegmen} The aim of this
paper is to obtain $C$ for a Hamiltonian system with $n$ degrees
of freedom and $m$ constants of motion $C_1,\cdots ,C_m$, with
$m\geq 2n$.

First, we begin by reviewing the basic features of the Nambu
formalism. Let $M$ be an $ND$ smooth manifold and let $C^{\infty
}(M)$ be the linear space of smooth real-valued functions on $M$.
The real multilinear map
\begin{equation}
\{,...,\}:\underbrace{C^{\infty}(M)\times \cdots \times
C^{\infty}(M)}_{N\;times}\rightarrow C^{\infty}(M)
\end{equation}
defines NB of $N$-th order satisfying the properties
skew-symmetry, Leibniz rule and generalized Jacobi identity
(fundamental identity). \cite{Takhtajan} When one considers this
NB structure $C^{\infty }(M)$ admits another algebra structure.

The time evolution of $f \in C^{\infty}(M)$ is determined by $N-1$
Hamiltonian functions $C_{1},\dots ,C_{N-1}\in C^{\infty}(M)$ and
is described by the Nambu-Hamilton (NH) equations of motion
\begin{equation}\label{NH}
\frac{df}{dt}={\bf X}_{NH}(f)=\{f, C_{1},\dots, C_{N-1} \},
\end{equation}
where ${\bf X}_{NH}$ is called the NH vector field corresponding
to $C_{1},\dots ,C_{N-1}$.

Finally, we recall that when $M$ is a symplectic manifold of
dimension $N=2n$, $C^{\infty}(M)$ has also an infinite-dimensional
Lie algebra structure defined with respect to the Poisson bracket
(PB)
\begin{equation}
\{f,\;g\}_P=\sum_{j=1}^{n}\left( \partial _{q_{j}}f\partial
_{p_{j}}g-\partial _{p_{j}}f\partial _{q_{j}}g \right) ,
\end{equation}
where $({\bf q},\;{\bf p})=(q_{1},\dots ,\;q_{n},\;p_{1}, \dots
,\;p_{n})$ are the local Darboux canonical coordinates.

\section{Canonical Nambu Bracket and its Decomposition}

A concrete realization of NB was embodied with the following form
\begin{equation}\label{J}
\{f_{1},\dots ,f_{N}\}=\frac{\partial (f_{1},\dots
,f_{N})}{\partial (x_{1},\dots ,x_{N})}
\end{equation}
by Y. Nambu in the case of $M=\mathbf{R}^{N}$ and it is called the
canonical
NB. In (\ref{J}) ${\bf x}=(x_{1},\dots ,x_{N})$ denotes the local coordinates of $%
\mathbf{R}^{N}$ and the right hand side stands for the Jacobian of
the mapping $f=(f_{1},\dots ,f_{N}):\mathbf{R}^{N}\rightarrow
\mathbf{R}^{N}$.

For the systems with $n$ degrees of freedom ,i.e.,
$2n$-dimensional phase space, the canonical Nambu bracket is
determined explicitly by
\begin{equation}
\{f_{1},\dots ,f_{2n}\}=\frac{\partial (f_{1},\dots
,f_{2n})}{\partial (q_{1},p_{1},\dots ,q_{n},p_{n})}=
\epsilon_{i_{1}\cdots i_{2n}}\partial_{i_1}f_{1}\cdots
\partial_{i_{2n}}f_{2n}
\end{equation}
and it can be decomposed into a skew-symmetric product of PBs
which is a useful identity
\begin{equation}\label{DG}
\{f_{1},\dots,f_{2n}\}=\{ f_{1},f_{[2} \}_P\{ f_{3},\dots ,f_{2n]}
\},
\end{equation}
where the bracket $[\;]$ indicates the cyclic sum with respect to
its elements.\cite{Curtright} (Here and hereafter we will use the
Einstein summation convention for all repeated indices.)

For coordinate-free expression of the canonical NB we associate the $(N-1)$%
-form
\begin{equation}\label{form}
\Gamma ={\rm d}C_{1}\wedge \cdots \wedge {\rm d}C_{N-1}
\end{equation}
to given $N-1$ Hamiltonian functions $C_{i}$. In (\ref{form}), $%
{\rm d}$ and $\wedge $ denote the usual exterior derivative and
exterior product
of Cartan calculus. If we employ the Hodge map $^{\star }:\Lambda ^{p}(%
\mathbf{R}^{N})\rightarrow \Lambda ^{N-p}(\mathbf{R}^{N})$ between
the space of $p$-forms $\Lambda ^{p}(\mathbf{R}^{N}),\;p\leq N$
and the space of $(N-p)$-forms $\Lambda^{N-p}(\mathbf{R}^{N})$, an
easy calculation shows that
\begin{equation}\label{dfform}
^{\star}({\rm d}f\wedge\Gamma)=\{f, C_{1}, \dots , C_{N-1}\}= \frac{%
\partial(f,C_{1},\dots, C_{N-1})}{\partial(x_{1},\dots, x_{N})}.
\end{equation}

Finally, for the purpose of this study, we write the NH equation
for any $f$ as
\begin{equation}\label{cfdot}
^{\star }(df\wedge \Gamma )=C\dot{f},
\end{equation}
where $C$ is also a constant of motion. $C$ will be referred to as
normalization constant corresponding to the set $C_1,\dots
,C_{N-1}$ and will be specified from the requirement that
(\ref{cfdot}) produces the correct Hamiltonian equations of motion
in the case of $N=2n$. The requirement of nontrivial $C$ is
inevitable at least when $n\geq 2$. The NB given in (\ref{cfdot})
will be used in the rest of the text.
\section{Nambu Brackets and Constraint Functionals}

While some authors have investigated the connection of Nambu
dynamics to Dirac's constraint formalism,
\cite{Bayen,Lassig,Grabowski} our approach will depend on
functions of these constraints. We consider a system with $n$
degrees of freedom, i.e., a $2n$-dimensional phase space with
coordinates $(q_{1},p_{1},\dots ,q_{n},p_{n}),\quad n\geq 2$. We
suppose that the system acquires $m\geq 2n$ constants of motion
$C_{1},\dots ,C_{m}$ constituting a polynomial algebra closing in
PB. Though these restrictions seem a bit rigorous, it may be
remarkable to emphasize that most of the examples in the
literature are of this kind. We impose no restriction on
involutive properties of the constants of motion. In this sense,
the system need not be integrable or superintegrable. For such a
system, when possible, the number of functionally independent
constants of motion is $2n-1$ and therefore there are independent
$m-(2n-1)=s$ functional relations
\begin{equation}
F_{j}=F_{j}(C_{1},\dots ,C_{m})=0,\qquad (j=1,\dots ,s)
\end{equation}
between $C_{i}$s \cite{Kaplan}. We will call the $F_j$s constraint
functionals to avoid any confusion. In some cases, too many
constraint functionals may appear, but any independent $s$ of them
are enough to operate the formalism, hence the choice of the
constraints for these systems is not unique. We divide the
derivation into two main cases to obtain a full treatment. First,
for the case $s\leq 2n-1$, let us choose the independent set of
$F_1,\dots ,F_s$ and construct the $s$-form
\begin{equation}\label{alpha}
\alpha = dF_1\wedge\cdots\wedge dF_s= \frac{\partial F_1}{\partial
C_{i_1}}\cdots \frac{\partial F_s}{\partial C_{i_s}}
\;dC_{i_1}\wedge\cdots\wedge dC_{i_s}.
\end{equation}
Now we take the arbitrary set $C_{i_{s+1}},\dots ,C_{i_{2n-1}}$
and construct the $((2n-1)-s)$-form
\begin{equation}\label{beta}
\beta = dC_{i_{s+1}}\wedge\cdots\wedge dC_{i_{2n-1}},
\end{equation}
and then the $(2n-1)$-form
\begin{eqnarray}\label{ab}
\alpha \wedge \beta &=& dF_1\wedge\cdots\wedge dF_s\wedge
dC_{i_{s+1}}\wedge\cdots\wedge dC_{i_{2n-1}}\nonumber \\
&=&\frac{\partial F_1}{\partial C_{i_1}}\cdots \frac{\partial
F_s}{\partial C_{i_s}} \;dC_{i_1}\wedge\cdots\wedge dC_{i_s}\wedge
dC_{i_{s+1}} \wedge \cdots \wedge dC_{i_{2n-1}}.
\end{eqnarray}
When we multiply (\ref{ab}) by $df$ and apply the Hodge map, we
get a $2n$-th order NB
\begin{equation}
^\star (df\wedge\alpha\wedge\beta )=\{f,F_1,\dots
,F_s,C_{i_{s+1}}, \dots ,C_{i_{2n-1}}\}.
\end{equation}
Obviously, if the $F_j$s are written in terms of the phase space
coordinates then they get zero, therefore
\begin{equation}\label{zero}
^\star (df\wedge\alpha\wedge\beta )= \frac{\partial F_1}{\partial
C_{i_1}}\cdots \frac{\partial F_s}{\partial C_{i_s}} \; ^\star
(df\wedge dC_{i_1}\wedge\cdots\wedge dC_{i_{2n-1}})=0.
\end{equation}
Each NB in \ref{zero}, by the definition (\ref{cfdot}), implies a
normalization constant associated with $C_{i_1},\dots
,C_{i_{2n-1}}$, i.e.,
\begin{equation}\label{dotf}
^\star (df\wedge dC_{i_1}\wedge\cdots\wedge dC_{i_{2n-1}})=
\dot{f}C_{i_1\cdots i_{2n-1}}.
\end{equation}
Thus this leads to a linear homogeneous system
\begin{equation}\label{heq}
\frac{\partial F_1}{\partial C_{i_1}}\cdots \frac{\partial
F_s}{\partial C_{i_s}} \;C_{i_1\cdots i_{2n-1}}=0
\end{equation}
including $\left( \begin{array}{c}m\\2n-1-s\end{array}\right)$
equations. It is easy to see by the basic solution techniques that
(\ref{heq}) has infinitely many solutions for the unknowns
$C_{i_1\cdots i_{2n-1}}$. But we require a solution which is
nontrivial and compatible with the decomposition (\ref{DG}). The
suitable solution can be chosen as
\begin{equation}\label{coef}
C_{i_1\cdots i_{2n-1}}=\epsilon_{i_1\cdots i_{2n-1}i_{2n}\cdots
i_m} \;\frac{\partial F_1}{\partial C_{i_{2n}}}\cdots
\frac{\partial F_s}{\partial C_{i_m}} =\pm \frac{\partial
(F_1,\dots ,F_s)} {\partial (C_{i_{2n}},\dots ,C_{i_m})},
\end{equation}
where the sign $\pm$ is determined by the Levi-Civita tensor in
the second term. Indeed, this choice argues with (\ref{heq}),
\begin{eqnarray}
& &\epsilon_{i_1\cdots i_s\underbrace{i_{s+1}\cdots
i_{2n-1}}_{(2n-1-s)\;indices} i_{2n}\cdots i_m}\;\frac{\partial
F_1}{\partial C_{i_1}}\cdots \frac{\partial F_s} {\partial
C_{i_s}} \frac{\partial F_1}{\partial C_{i_{2n}}}\cdots
\frac{\partial F_s}{\partial C_{i_m}} \nonumber \\
&=&(-1)^{s(2n-1-s)}\epsilon_{i_{s+1}\cdots i_{2n-1}i_1\cdots i_s
i_{2n}\cdots i_m}\frac{\partial F_1}{\partial C_{i_1}}\cdots
\frac{\partial F_s}{\partial C_{i_s}} \frac{\partial F_1}{\partial
C_{i_{2n}}}\cdots
\frac{\partial F_s}{\partial C_{i_m}} \nonumber \\
&=&\pm \frac{\partial (F_1,\dots ,F_s,F_1,\dots ,F_s)} {\partial
(C_{i_1},\dots ,C_{i_s},C_{i_{2n}},\dots ,C_{i_m})}=0.
\end{eqnarray}
An illustrative example may be convenient to be more explicit.
Consider the system with two degrees of freedom including the
constants of motion $C_1,\dots ,C_5$, which is also considered in
Subsec.4.1.. Thus there exist two independent constraint
functionals, $F_1$, $F_2$. Therefore
\begin{eqnarray}
\alpha =dF_1\wedge dF_2 &=&\frac{\partial F_1}{\partial C_{i_1}}
\frac{\partial F_2}{\partial C_{i_2}}\;dC_{i_1}\wedge
dC_{i_2}\nonumber \\
&=&\frac{\partial (F_1,F_2)}{\partial (C_{i_1},C_{i_2})}
\;dC_{i_1}\wedge dC_{i_2},\quad (i_1 < i_2);
\end{eqnarray}
and $\beta =dC_{i_3}$. For an arbitrary $C_{i_3}$, say $C_1$, the
condition $^\star (df\wedge dF_1\wedge dF_2\wedge dC_1)=0$ implies
\begin{eqnarray}
& &\frac{\partial (F_1,F_2)}{\partial (C_2,C_3)}\;C_{231}
+\frac{\partial (F_1,F_2)}{\partial (C_2,C_4)}\;C_{241}
+\frac{\partial (F_1,F_2)}{\partial (C_2,C_5)}\;C_{251} \nonumber \\
&+&\frac{\partial (F_1,F_2)}{\partial (C_3,C_4)}\;C_{341}
+\frac{\partial (F_1,F_2)}{\partial (C_3,C_5)}\;C_{351}
+\frac{\partial (F_1,F_2)}{\partial (C_4,C_5)}\;C_{451}=0.
\end{eqnarray}
Referring back to (\ref{coef}), one computes easily
\begin{eqnarray}
& &2\frac{\partial (F_1,F_2)}{\partial (C_2,C_3)} \frac{\partial
(F_1,F_2)}{\partial (C_4,C_5)} -2\frac{\partial
(F_1,F_2)}{\partial (C_2,C_4)} \frac{\partial (F_1,F_2)}{\partial
(C_3,C_5)} +2\frac{\partial (F_1,F_2)}{\partial (C_2,C_5)}
\frac{\partial (F_1,F_2)}{\partial (C_3,C_4)}\nonumber \\
& &= \frac{\partial (F_1,F_2,F_1,F_2)} {\partial
(C_2,C_3,C_4,C_5)}=0.
\end{eqnarray}

At the second stage, i.e., for the case $s>2n-1$, we choose any
$2n-1$ constraints from the set $F_1,\dots ,F_s$. With the same
argument followed in the previous case, it is easy to construct
the $2n$-form
\begin{eqnarray}
& &^\star (df\wedge dF_{j_1}\wedge \cdots
\wedge dF_{j_{2n-1}})\nonumber \\
& &=\frac{\partial F_{j_1}}{\partial C_{i_1}}\cdots \frac{\partial
F_{j_{2n-1}}}{\partial C_{i_{2n-1}}} \;^\star (df\wedge
dC_{i_1}\wedge \cdots \wedge dC_{i_{2n-1}})=0
\end{eqnarray}
generating the system with $\left(
\begin{array}{c}s\\2n-1\end{array}\right)$ equations
\begin{equation}\label{sheq}
\frac{\partial F_{j_1}}{\partial C_{i_1}}\cdots \frac{\partial
F_{j_{2n-1}}}{\partial C_{i_{2n-1}}} \;C_{i_1\cdots i_{2n-1}}=0.
\end{equation}
Since the choice of the set $F_{j_1},\dots ,F_{j_{2n-1}}$ is
arbitrary, (\ref{coef}) is always a solution to (\ref{sheq}):
\begin{eqnarray}
& &\epsilon_{i_1\cdots i_m}\;\frac{\partial F_{j_1}}{\partial
C_{i_1}} \cdots \frac{\partial F_{j_{2n-1}}}{\partial
C_{i_{2n-1}}} \frac{\partial F_1}{\partial C_{i_{2n}}}\cdots
\frac{\partial F_s}{\partial C_{i_m}}\nonumber \\
& &=\pm \frac{\partial (F_{j_1},\dots ,F_{j_{2n-1}},F_1,\dots
,F_s)} {\partial (C_{i_1},\dots ,C_{i_{2n-1}},C_{i_{2n}},\dots
,C_{i_m})}=0.
\end{eqnarray}
Consequently, if (\ref{dotf}) is recombined with (\ref{coef}), one
concludes that
\begin{eqnarray}\label{final}
^{\star }(df\wedge dC_{i_1}\wedge \cdots \wedge dC_{i_{2n-1}})
&=&\{f,C_{i_1},\dots ,C_{i_{2n-1}}\}\nonumber \\
&=&\epsilon_{i_1\cdots i_{2n-1}i_{2n}\cdots i_m} \;\frac{\partial
F_1}{\partial C_{i_{2n}}}\cdots
\frac{\partial F_s}{\partial C_{i_m}}\dot{f}\nonumber \\
&=&\pm \frac{\partial (F_1,\dots ,F_s)} {\partial
(C_{i_{2n}},\dots ,C_{i_m})}\dot{f}.
\end{eqnarray}
The formalism (\ref{final}) possesses several NBs all are in
accordance with correct equations of motion. It may be remarkable
to emphasize that we have no any restriction about the
independence of the set $C_{i_{2n}},\dots ,C_{i_m}$. Thus, even if
they are not independent, this does not destroy the validity of
the formalism. On the other hand, the general result (\ref{final})
also justifies the statement: If a NB includes any dependent
subset of the constants of motion, then it vanishes. We shall make
these remarks more precise in the discussion of the examples.

Now, as a corollary, we conclude the statement: If one of the
constants of motion, say $C_k$, is taken as the Hamiltonian, then
the decomposition (\ref{DG}) can, by discarding the Hamiltonian,
be written in terms of the constraint functionals. The proof is
straightforward: By the decomposition (\ref{DG}),
\begin{equation}\label{decompH}
\{f,C_k,C_{i_1},\dots ,C_{i_{2n-2}}\}
=\{f,C_k\}_{P}\{C_{i_1},\dots ,C_{i_{2n-2}}\}
=\dot{f}\{C_{i_1},\dots ,C_{i_{2n-2}}\},
\end{equation}
on the other hand, by (\ref{final}),
\begin{equation}
\{f,C_k,C_{i_1},\dots ,C_{i_{2n-2}}\}= \epsilon_{ki_1\cdots
i_{m-1}} \;\frac{\partial F_1}{\partial C_{i_{2n-1}}}\cdots
\frac{\partial F_s}{\partial C_{i_{m-1}}}\dot{f}\;,
\end{equation}
thus
\begin{equation}\label{decomp}
\{C_{i_1},\dots ,C_{i_{2n-2}}\} =\pm \frac{\partial (F_1,\dots
,F_s)} {\partial (C_{i_{2n-1}},\dots ,C_{i_{m-1}})}.
\end{equation}
In particular, for the case $n=2$, (\ref{decomp}) holds for the
PBs.

Finally, we talk about determination of the $F_j$s. Note that, as
a general result of (\ref{DG}), for the NBs including the
Hamiltonian, the normalization constant $C$ is obtained easily in
terms of the PBs. Therefore this observation supplies a helpful
guide in determining the constraint functionals. Under this
circumstance, $s-1$ functional relations can be taken as the
constraint functionals without any rearrangement. Thus, the
construction of the last one is reduced to the problem of finding
a function whose derivatives are known. Since the PBs of the
constants of motion close to constitute a polynomial algebra, this
gives us an easy integration process.

Having shown how to construct such a formalism we will give
examples to confirm its correctness. To be more clear, the first
example (harmonic oscillator) has been analyzed in detail. The
examples have been chosen in a variety so that they include
various alternatives when considering the dimension of the phase
space and the number of the constraint functionals. In all
examples, although the construction of constants of motion is not
unique, we kept the forms and numbers of them just as appeared in
the literature cited in the text.

\section{Systems with Two and Three Degrees of Freedom}

\subsection{Harmonic oscillator}
Although we take the system as the one with two degrees of
freedom, its $2n,\;(n\geq 2)$ dimensional extension is obtainable
as a special case from the Winternitz system given in Subsec.5.1.
The system is described by the Hamiltonian
\begin{equation}
C_1=H={\bf p}^{2}/2+k{\bf q}^{2}/2,
\end{equation}
where ${\bf q}^{2}={q_1}^2+q_{2}^2$ and ${\bf
p}^{2}={p_1}^2+p_{2}^2$ \cite{Chatterjee}. Suppose that in
addition to the Hamiltonian we are given the following set of
equations as the constants of motion,
\begin{eqnarray}
C_2&=&p_1^2/2+kq_1^2/2,\quad C_3=p_2^2/2+
kq_2^2/2,\nonumber \\
C_4&=&q_1p_2-q_2p_1,\quad C_5=p_1p_2+kq_1q_2.
\end{eqnarray}
Their nonvanishing PBs are given by
\begin{eqnarray}
\{C_2,C_4\}_P&=&-\{C_3,C_4\}_P=-C_5,\nonumber \\
\{C_2,C_5\}_P&=&-\{C_3,C_5\}_P=kC_4,\nonumber \\
\{C_4,C_5\}_P&=&-2(C_2-C_3).
\end{eqnarray}
If the functional relation $C_1=C_2+C_3$ is taken as the first
constraint such that $F_1=C_1-C_2-C_3$, then the other which is
compatible with the all PBs via (\ref{decomp}) can be constructed
as the following
\begin{equation}
F_2=2C_2C_3-\frac{1}{2}kC_4^2-\frac{1}{2}C_5^2
\end{equation}
so that
\begin{eqnarray}
\{C_2,C_4\}_P&=&-\frac{\partial (F_1,F_2)}{\partial (C_3,C_5)}
=\frac{\partial F_2}{\partial C_5}=-C_5, \nonumber \\
\{C_2,C_5\}_P&=&\frac{\partial (F_1,F_2)}{\partial (C_3,C_4)}
=-\frac{\partial F_2}{\partial C_4}=kC_4, \nonumber \\
\{C_4,C_5\}_P&=&\frac{\partial (F_1,F_2)}{\partial (C_2,C_3)}
=-\frac{\partial F_2}{\partial C_3} +\frac{\partial F_2}{\partial
C_2}=-2(C_2-C_3).
\end{eqnarray}
Now we list two of the NBs as the sample,
\begin{eqnarray}
\{f,C_1,C_2,C_4\}&=&-\frac{\partial (F_1,F_2)}
{\partial (C_3,C_5)}\dot{f}=-C_5\dot{f},\nonumber \\
\{f,C_1,C_2,C_3\}&=& \frac{\partial (F_1,F_2)}{\partial (C_4,C_5)}
\dot{f}=0
\end{eqnarray}
which are consistent with
\begin{eqnarray}
\{f,C_1,C_2,C_4\}&=&\{f,C_1\}_P\{C_2,C_4\}_P=-C_5\dot{f},\nonumber \\
\{f,C_1,C_2,C_3\}&=&\{f,C_1\}_P\{C_2,C_3\}_P=0.
\end{eqnarray}
On the other hand, for the NBs not including the Hamiltonian, $C$
is not so evident via (\ref{DG}). For example
\begin{eqnarray}
\{f,C_3,C_4,C_5\}&=&\{f,C_3\}_P\{C_4,C_5\}_P+
\{f,C_5\}_P\{C_3,C_4\}_P+\{f,C_4\}_P\{C_5,C_3\}_P \nonumber \\
&=&-2(C_2-C_3)\{f,C_3\}_P+C_5\{f,C_5\}_P+kC_4\{f,C_4\}_P,
\end{eqnarray}
but it is straightforward to obtain it by the virtue of
\begin{equation}
\{f,C_3,C_4,C_5\}=\frac{\partial (F_1,F_2)}{\partial (C_1,C_2)}
\dot{f}=2C_3\dot{f}.
\end{equation}

If we turn off the Hamiltonian, the set $C_2,C_3,C_4,C_5$ is still
closed in PB and we need only one constraint functional, $F=-F_2$,
thus
\begin{equation}
\{f,C_3,C_4,C_5\}=-\frac{\partial F}{\partial C_2}\dot{f}
=2C_3\dot{f}.
\end{equation}

Now consider the bracket
\begin{equation}
\{f,C_1,C_4,C_5\}=\frac{\partial (F_1,F_2)}{\partial (C_2,C_3)}
\dot{f}=-2(C_2-C_3)\dot{f},
\end{equation}
and take artificially $C_6=C_2-C_3$ so that $C_2,C_3$ and $C_6$
are not independent. Again
\begin{equation}
\{f,C_1,C_4,C_5\}=\frac{\partial (F_1,F_2,F_3)}{\partial
(C_6,C_2,C_3)} \dot{f}=-2(C_2-C_3)\dot{f},
\end{equation}
where $F_3=C_6-C_2+C_3$.

Finally, for the case $s>2n-1$, let $C_7=C_4C_5$. Of much NBs, we
have chosen
\begin{equation}
\{f,C_3,C_4,C_5\}=\frac{\partial (F_1,F_2,F_3,F_4)} {\partial
(C_6,C_7,C_1,C_2)}=2C_3\dot{f},
\end{equation}
where $F_4=C_7-C_4C_5$.

\subsection{Smorodinsky-Winternitz system}

Smorodinsky-Winternitz system consists of a set of four
Hamiltonians which have potential form, i.e., $H={\bf
p}^{2}/{2}+V$ \cite{Winternitz}. All potentials are separable into
at least two coordinate systems and they also admit
superintegrable structure. We will consider symbolically only the
system
\begin{equation}
C_1=H={\bf p}^{2}/{2}+\omega^2 (4{q_1}^2 +{q_2}^2)+\alpha_1 q_1
+\alpha_2 /{q_2}^2,
\end{equation}
where all Greek letters are some real constants, $q =
(q_1^{2}+q_2^{2})^{1/2}$ and ${\bf p}^2={p_1}^2+p_{2}^2$. All
constants of motions are at most quadratic in momenta,
\begin{eqnarray}
C_2&=&{p_1}^2/2+\alpha_1 q_1 +4\omega^2{q_1}^2,\nonumber  \\
C_3&=&2p_2L_3-4\omega^2 q_1 q_2^2 +
4\alpha_2 q_1 /{q_2^2}-\alpha_1 q_2^2,\nonumber \\
C_4&=&-2(\alpha_1 +8\omega^2 q_1 )q_2 p_2 -p_1 (2p_2^2 -4\omega^2
q_2^2 + 4\alpha_2/{q_2^2}),
\end{eqnarray}
where $L_3 = q_1 p_2 -q_2 p_1$ is the third component of angular
momentum. Their nonvanishing PBs close in a Poisson algebra
\begin{eqnarray}\label{SWPB}
\{ C_2, C_3\}_P&=&C_4, \nonumber\\
\{ C_2, C_4\}_P&=&
4\alpha_1 C_2-8\omega^2 C_3-4\alpha_1 C_1, \nonumber\\
\{ C_3, C_4\}_P&=&-48C_2^2+64C_1 C_2-4\alpha_1C_3+
64\omega^2\alpha_2-16 C_1^2,
\end{eqnarray}
admitting a Casimir
\begin{eqnarray}\label{SWC}
F&=&C_4^2 /2 -4\alpha_1 C_2 C_3 +4\omega^2 C_3^2 +4\alpha_1 C_1
C_3 -16C_2^3 \nonumber \\
&+& 32C_1 C_2^2 +64\omega^2 \alpha_2 C_2 -16 C_1^2 C_2
+4\alpha_1^2 \alpha_2
\end{eqnarray}
as the constraint functional \cite{Daskaloyannis}. Note that the
constraint (\ref{SWC}) can be obtained as the integration of the
PBs in (\ref{SWPB}). This observation justifies the expressions
\begin{equation}
\{ C_2, C_3\}_P=\frac{\partial F}{\partial C_4},\quad \{ C_2,
C_4\}_P=-\frac{\partial F}{\partial C_3},\quad \{ C_3,
C_4\}_P=\frac{\partial F}{\partial C_2}.
\end{equation}
And hence two of possible NBs are
\begin{eqnarray}
\{f,C_1,C_2,C_4\}&=&-\frac{\partial F}{\partial C_3}
\dot{f}=-(4\alpha_1 C_2-8\omega^2 C_3-4\alpha_1 C_1)\dot{f}, \nonumber \\
\{f,C_2,C_3,C_4\}&=&-\frac{\partial F}{\partial C_1}
\dot{f}=-(4\alpha_1C_3+32C_2^2-32C_1C_2)\dot{f}.
\end{eqnarray}

\subsection{Kepler-Coulomb system}

Let us concentrate now on the 6D Kepler-Coulomb Hamiltonian
\begin{equation}
H={\bf p}^{2}/2-\alpha/q,
\end{equation}
where $\alpha$ is a real constant, ${\bf
p}^{2}={p_1}^2+p_{2}^2+{p_3}^2$ and
$q=({q_1}^2+q_{2}^2+{q_3}^2)^{1/2}$ \cite{Chatterjee}. Because of
the rotational symmetry of the system, the angular momentum ${\bf
L}$ is integral invariant and hence its components can be taken as
constants of motion, i.e.,
\begin{equation}
L_1=q_2p_3-q_3p_2, \quad L_2=q_3p_1-q_1p_3, \quad
L_3=q_1p_2-q_2p_1.
\end{equation}
(From now on, we will write the constants of motion just as
appeared in the system without corresponding any $C_i$ to them to
avoid any confusion in mind and to keep their symmetries in
writting). Moreover, there also exists an extra invariant arising
from that the particle has a closed orbit. This invariant is
called the Runge-Lenz vector ${\bf A}$ given by
\begin{equation}
{\bf A}={\bf p}\times{\bf L}-\alpha {\bf q}/q.
\end{equation}
In addition to the previous constants of motion,
\begin{eqnarray}
A_1&=p_2L_3-p_3L_2-\alpha q_1/q, \nonumber\\
A_2&=p_3L_1-p_1L_3-\alpha q_2/q,\nonumber\\
A_3&=p_1L_2-p_2L_1-\alpha q_3/q.
\end{eqnarray}
Consequently we have seven constants of motion satisfying the
commutations
\begin{equation}
\{L_a,L_b\}_P=\epsilon_{abc}L_c,\quad
\{A_a,A_b\}_P=-2H\epsilon_{abc}L_c,\quad
\{L_a,A_b\}_P=\epsilon_{abc}A_c,
\end{equation}
and the following functional relations
\begin{equation}
{\bf A}\cdot {\bf L}=0,\qquad A^2=1+2HL^2
\end{equation}
which are candidates for the constraint functionals. Thus if we
choose these functions as follows
\begin{eqnarray}
F_1&=&A_1L_1+A_2L_2+A_3L_3,\nonumber\\
F_2&=&\frac{1}{2}+H(L_1^2+L_2^2+L_3^2)-
\frac{1}{2}(A_1^2+A_2^2+A_3^2),
\end{eqnarray}
(\ref{decomp}) implies several brackets such as
\begin{equation}
\{L_1,L_2,L_3,A_1\}=\frac{\partial (F_1,F_2)}{\partial (A_2,A_3)}=
A_2L_3-A_3L_2.
\end{equation}
This is compatible with the decomposition
\begin{eqnarray}
\{L_1,L_2,L_3,A_1\}&=&\{L_1,L_2\}_P\{L_3,A_1\}_P+
\{L_1,A_1\}_P\{L_2,L_3\}_P \nonumber \\
&+&\{L_1,L_3\}_P\{A_1,L_2\}_P.
\end{eqnarray}
So, one of the NBs is
\begin{eqnarray}
\{f,H,L_1,L_2,L_3,A_1\}&=&\{f,H\}_P \{L_1,L_2, L_3, A_1\}\nonumber\\
&=&\frac{\partial (F_1,F_2)}{\partial
(A_2,A_3)}\dot{f}=(A_2L_3-A_3L_2)\dot{f}.
\end{eqnarray}
Other two of them may be chosen as the following
\begin{eqnarray}
\{f,L_1,L_2,L_3,A_1,A_2\}&=& \frac{\partial (F_1,F_2)}{\partial
(A_3,H)}\dot{f}=L_3({L_1}^2+{L_2}^2+{L_3}^2)\dot{f},\nonumber\\
\{f,H,L_1,L_2,A_2,A_3\}&=& \frac{\partial (F_1,F_2)}{\partial
(L_3,A_1)}\dot{f}=-(A_1A_3+2HL_1L_3)\dot{f}.
\end{eqnarray}

\section{Systems with $n$ Degrees of Freedom}
\subsection{Winternitz system}
Winternitz system is the arbitrary dimensional generalization of
one of the Smorodinsky - Winternitz Hamiltonians mentioned above.
Their constants of motion are constructed by using the Lax matrix
representation \cite{Evans}. The particle's Hamiltonian in $n$
degrees of freedom is given by
\begin{equation}
H=\frac{1}{2}\sum_{i=1}^{n}(p_i^{2}+k^{2}x_i^{2}+
\frac{k_i^2}{x_i^2}),
\end{equation}
where $k$ and the $k_i$ are real constants. (Throughout this
section all subscripts range from 1 to $n$). The elements of
degeneracy group $SU(n)$ of the Winternitz system are taken as the
constants of motion all commuting with the Hamiltonian. First
group of these $n^2$ functions has the form
\begin{equation}\label{tii}
T_{ii}=\frac{1}{2k}(H_{i}-kk_{i}),
\end{equation}
where the conserved quantity $H_i$ is taken as the energy in the
$i$-th direction. Second group is given by the ansatz
\begin{equation}
T_{ij}=f(H_{i})f(H_{j})A_{i}A_j^{\ast},\quad i\neq j,
\end{equation}
where
\begin{equation}
f(H_{i})=\left( \frac{2k}{H_{i}+kk_i}\right)^{1/2},
\end{equation}
and
\begin{equation}
A_{i}=\frac{1}{4k}\left( p_i^{2}+\frac{k_i^2}{x_i^2}-
k^{2}x_i^{2}+2{\rm i}kx_{i}p_{i}\right).
\end{equation}
The PBs of the functions $T_{ij}$ argue
\begin{equation}
\{ T_{ij},T_{rs}\}_{P} ={\rm i}\delta_{jr}T_{is}- {\rm
i}\delta_{is}T_{rj}.
\end{equation}
With this argument, there exist totally $n^2+1$ constants of
motion and there must be functional relations between the
invariants. One of the functional relations, so by (\ref{tii}), is
the simplest one
\begin{equation}
H=\sum_{i=1}^{n}(2kT_{ii}+kk_{i}).
\end{equation}
After a series of calculations, the others can be expressed as
\begin{equation}
T_{ij}T_{jk}=T_{jj}T_{ik}.
\end{equation}
Note, by referring to the discussion in Sec.3, that the number of
independent constraint functionals is $n^2-2n+2$.

After having been defined the Winternitz system and its
invariants, we now perform the case of three degrees of freedom as
an example. Despite the fact that it is possible to study with ten
constants of motion and five constraint functionals (or ten
functional relations), for the sake of simplicity, we prefer the
set of constants of motion $H,T_{11},T_{22},T_{33},T_{12},T_{13}$
which is also closed in the PB. The suitable choice for the only
one constraint functional is
\begin{equation}
F=T_{12}T_{13}\left[-\frac{H}{2k}+T_{11}+T_{22}+T_{33}+\frac{1}{2}(k_1+k_2+k_3)\right].
\end{equation}
Consequently among the all possible four nonvanishing NBs two of
them are listed in the following
\begin{eqnarray}
\{f,H,T_{11},T_{22},T_{12},T_{13}\}&=& \frac{\partial F}{\partial
T_{33}}\dot{f}=
T_{12}T_{13}\dot{f},\nonumber \\
\{f,T_{11},T_{22},T_{33},T_{12},T_{13}\}&=&- \frac{\partial
F}{\partial H}\dot{f}= \frac{1}{2k}T_{12}T_{13}\dot{f}.
\end{eqnarray}

\subsection{Free particle on $n$-sphere}

Our last example is a free particle moving on the surface of an
$n$-sphere with the radius $q=({q_1}^2+\cdots +{q_n}^2)^{1/2}$
\cite{Curtright}. The PB Lie algebra of the integral invariants
(charges of $so(n+1)$) is generated by the $n(n-1)/2$ rotation
elements $L_{\alpha \beta}=q_{\alpha}p_{\beta}-
q_{\beta}p_{\alpha},\;\alpha, \beta =1,\dots n$, and the $n$
momenta $P_\alpha=(1-q^2)^{1/2}p_\alpha$. The PBs of the
invariants are
\begin{equation}
\{ L_{\alpha\beta}, L_{\gamma\xi}\}_P
=\delta_{\beta\xi}L_{\alpha\gamma}+
\delta_{\alpha\gamma}L_{\beta\xi}-\delta_{\beta\gamma}L_{\alpha\xi}-
\delta_{\alpha\xi}L_{\beta\gamma},
\end{equation}
\begin{equation}
\{ P_\alpha, P_\beta \}_P=L_{\alpha\beta},\qquad \{
L_{\alpha\beta}, P_\gamma \}_P=\delta_{\alpha\gamma}P_\beta -
\delta_{\beta\gamma}P_\alpha .
\end{equation}
The Hamiltonian of the particle is given by
\begin{equation}\label{HS}
H=\frac{1}{2}(P_\alpha P_\alpha +L_{\beta\gamma}L_{\beta\gamma}),
\quad \beta < \gamma .
\end{equation}
Unlike the previous example, the algebra is closed by all of the
$n(n+1)/2+1$ constants of motion, and then we need $n(n-3)/2+2$
independent functionals to proceed the formalism. For this aim we
will use the following concluded functional relations as the
source of the constraint functionals;
\begin{eqnarray}\label{group1}
L_{[\alpha_1 \alpha_2}P_{\alpha_3 ]}&=0,\nonumber \\
L_{[\alpha_1 \alpha_2}P_{\alpha_3}P_{\alpha_4]}&=0,\nonumber \\
\qquad \qquad & \vdots  \nonumber  \\
L_{[\alpha_1 \alpha_2}P_{\alpha_3}P_{\alpha_4}\cdots
P_{\alpha_n]}&=0.
\end{eqnarray}
Additionally a second group appears as
\begin{equation}\label{group2}
L_{\alpha_1 [\alpha_2}L_{\alpha_3 \alpha_4 ]}=0.
\end{equation}
For each of the relations in (\ref{group1}) and (\ref{group2}),
the $\alpha_i$s can be chosen freely provided they are all
different from one another. It is clear to see that the number of
functional relations listed above exceeds the needed one too much,
but any set of independent, suitable rearranged $n(n-3)/2+2$
functionals does work. As was the case for the previous example,
we restricted ourselves to a particular, say $n=4$, degrees of
freedom. In that case, we have 11 constants of motion and
therefore we need four independent constraints. First one is the
simplest one, i.e., (\ref{HS}),
\begin{eqnarray}
F_1=H&-&\frac{1}{2}({P_1}^2+{P_2}^2+{P_3}^2+{P_4}^2)\nonumber \\
&-&\frac{1}{2}({L_{12}}^2+{L_{13}}^2+{L_{14}}^2+{L_{23}}^2+
{L_{24}}^2+{L_{34}}^2).
\end{eqnarray}
For the other three, a suitable choice can be written explicitly
as the following
\begin{eqnarray}
F_2&=&L_{12}L_{34}+L_{14}L_{23}-L_{13}L_{24}, \nonumber\\
F_3&=&L_{12}P_3-L_{13}P_2+L_{23}P_1,\nonumber \\
F_4&=&P_4-\frac{L_{14}}{L_{13}}P_3+ \frac{L_{34}}{L_{13}}P_1.
\end{eqnarray}

Now, as before, we choose a sample from the NBs,
\begin{eqnarray}
& &\{f,H,P_1,P_2,P_3,P_4,L_{12},L_{13}\} =\frac{\partial
(F_1,F_2,F_3,F_4)}{\partial
(L_{14},L_{23},L_{24},L_{34})}\dot{f}\nonumber \\
&=&\left[L_{12}P_2P_4+L_{13}P_3P_4-L_{14}({P_1}^2+
{P_2}^2+{P_3}^2)\right]\dot{f},
\end{eqnarray}
here we used the facts that $L_{[12}P_{3]}=0$, $L_{[12}P_{4]}=0$
and $L_{[13}P_{4]}=0$. As a further consequence, we remark that
another set
\begin{eqnarray}
{F_1}^{\prime}&=&F_1, \nonumber\\
{F_2}^{\prime}&=&L_{14}P_2P_3-L_{13}P_2P_4-L_{24}P_1P_3+
L_{23}P_1P_4, \nonumber \\
{F_3}^{\prime}&=&\frac{L_{23}}{P_3}-\frac{L_{24}}{P_4}+
\frac{L_{34}}{P_3P_4}P_2, \nonumber\\
{F_4}^{\prime}&=&\frac{P_3P_4}{P_2}L_{12}-P_3L_{14}+P_1L_{34}+
\frac{P_1P_4}{P_2}L_{23}
\end{eqnarray}
gives the same equations of motion. For the readers who may wonder
of other brackets, here we list two of them,
\begin{eqnarray}
\{f,P_1,P_2,P_3,P_4,L_{12},L_{13},L_{24}\}&=&-P_1P_2\dot{f},\nonumber\\
\{f,H,L_{12},L_{13},L_{14},L_{23},L_{24},L_{34}\}&=&0.
\end{eqnarray}

\section*{Acknowledgments}

The author wishes to express his appreciations to A. Ver{\c c}in
for his careful reading. This work was supported in part by the
Scientific and Technical Research Council of Turkey
(T\"{U}B\.{I}TAK).

\end{document}